# Superhydrophobic Sand Mulch Shifts Soil Evaporation from Temperature-Controlled to Diffusion-Limited Regimes

Amr Al-Zu'bi[1,2,3,#], Muhammad Subkhi Sadullah[1,2,3,#], Jiaqi Zheng[1,2,3], Lisa Exposito[1,2,3],

Adair Gallo Jr.[4], Himanshu Mishra[1,2,3*]

[1]Environmental Science and Engineering (EnSE) Program, Biological and Environmental Science and Engineering (BESE) Division, King Abdullah University of Science and Technology (KAUST), Thuwal, 23955-6900, Saudi Arabia.

[2]Center of Excellence for Sustainable Food Security, King Abdullah University of Science and Technology (KAUST), Thuwal 23955-6900, Kingdom of Saudi Arabia

[3]Water Desalination and Reuse Platform (WDRP), King Abdullah University of Science and Technology (KAUST), Thuwal 23955-6900, Kingdom of Saudi Arabia

[4]Terraxy LLC, King Abdullah University of Science and Technology (KAUST), Thuwal 23955-6900, Kingdom of Saudi Arabia

[#]Equal author contribution
[*]Corresponding author: himanshu.mishra@kaust.edu.sa

**Abstract:**

In hot arid and semi-arid regions, substantial irrigation water is lost through surface evaporation under intense solar irradiation and high temperatures, limiting freshwater sustainability and crop productivity. Superhydrophobic Sand (SHS) mulch—a plastic-free, bio-inspired technology—has been proposed as a dry diffusion barrier to suppress evaporative losses. Here, we combine controlled column experiments with heat and mass transfer modeling to quantify how SHS thickness and soil properties govern evaporation under fixed irradiation. Relative to unmulched controls, a 5 mm SHS layer reduced evaporative flux by 65% in fine sand and 63% in coarse sand, while a 10 mm layer reduced flux by 83% and 70%, respectively. Notably, soil-type trends reversed after mulching: although unmulched fine sand exhibited 37.5% higher evaporation than coarse sand, application of a 10 mm SHS layer reduced fine-sand evaporation to 40% below that of coarse sand. To explain this counterintuitive behavior, we developed a coupled heat and vapor transport model incorporating soil thermophysical properties and diffusion through the porous mulch layer. The model accurately predicted steady-state temperature profiles and evaporation rates for both mulched and unmulched systems. Our results show that SHS mulch shifts evaporation from a surface-temperature-controlled regime to a diffusion-limited regime governed by mulch thickness and soil thermal conductivity. This mechanistic understanding clarifies the performance of SHS and supports its potential to enhance irrigation efficiency in arid agricultural and landscaping applications.

# 1. Introduction

Global drylands cover approximately 46% of Earth's land surface and support billions of people under conditions of limited freshwater availability, high solar irradiation, and increasing aridity[1-4]. In many arid and semi-arid regions, up to 70% of annual freshwater resources are allocated to agriculture, yet a substantial fraction of applied irrigation water is lost through soil evaporation, particularly during early growth stages when canopy cover is minimal[5]. Food–water security is often intimately coupled with regional stability and is a matter of national and international significance[6-8].

While drip and subsurface irrigation reduce runoff and percolation losses by delivering water directly to the root zone[9], they do not eliminate surface evaporative losses driven by radiative heating and atmospheric demand[10]. Plastic mulches are widely used to suppress evaporation; however, their environmental persistence, disposal challenges, and microplastic contamination have raised sustainability concerns[11]. Moreover, under hot climates, plastic films can elevate root-zone temperatures, potentially affecting plant performance[11, 12]. Despite their widespread use, the coupled roles of soil and mulch thermophysical properties—such as albedo and thermal conductivity—in regulating evaporative flux and temperature distribution remain insufficiently quantified.

Inspired by the many water-repellent plants and animals in nature[13, 14], researchers have developed coating-free solutions for liquid–vapor separation[15], underwater gas restoration[16, 17], prolonging the shelf-life of stored seeds in engineered jute bags[18], and mitigating cavitation erosion of underwater machinery[19] and reducing frictional drag of water on surfaces through air entrapment[20]. Along these lines, a plastic-free mulching technology – Superhydrophobic Sand (SHS) – has been developed to bolster irrigation efficiency in arid lands[21]. SHS is manufactured by applying a nanoscale layer of biodegradable paraffin wax onto common desert sand grains, such that the paraffin-to-sand mass ratio is 1:500.[22] Consequently, the dual effect of paraffin's hydrophobicity and sand's surface roughness at the grain level (nano- and micrometer-scale) and macroscopic (cm-scale), gives rise to superhydrophobicity. As liquid water contacts SHS, air is trapped at the solid–water interface, and capillary pressure, arising due to liquid-vapor interfacial curvature, prevents water's spontaneous imbibition or wicking into the SHS layer. Consequently, when SHS mulch (e.g., 1 cm-thick layer) is applied on top of wet soil, it affords as a dry diffusion barrier, shielding liquid water from direct exposure to sunlight and wind. Multiyear field trials with tomato, wheat, and barley have demonstrated +70%, +40%, and +20% enhanced plant yield compared with fertilized and irrigated controls[21]. Yields of SHS-treated tomato plants in this trial and of sweet pepper plants (in another field trial) were found to be at par with plastic mulching[23]. Effects of SHS have also been studied under reduced irrigation conditions through pot studies with tomato[24] and moringa[25], and found to be significantly beneficial in terms of plant yield and biomass, respectively, under reduced irrigation conditions. Notably, SHS biodegrades in about a year under soil microbial pressure, integrating into the sandy soil matrix.

While agronomic benefits have been demonstrated, the fundamental physical mechanisms governing evaporation and temperature distribution in SHS–soil systems remain unclear. In particular:

(i) How do soil grain size, thermal conductivity, and albedo influence steady-state temperature profiles?

(ii) Does SHS reduce evaporation primarily by lowering surface temperature, or by introducing vapor diffusion resistance?

(iii) Why do different soil textures respond differently to SHS application?

Here, we investigate the coupled heat and mass transfer processes in SHS-mulched and unmulched sandy soils using controlled column experiments and analytical modeling. By systematically varying soil texture, mulch thickness, and water table depth under fixed irradiation, we quantify evaporative fluxes and temperature profiles. Complementary modeling reveals that unmulched evaporation is predominantly surface-temperature controlled, whereas SHS mulching shifts the system to a diffusion-limited regime governed by mulch thickness and soil thermal conductivity. These findings provide mechanistic understanding necessary for optimizing SHS performance in arid agricultural and landscaping systems.

## 2. Materials and Experimental Setup
### 2.1. Materials Characterization

Superhydrophobic sand (SHS) was produced by mixing common desert sand with paraffin wax, following a previously reported protocol[21]. Briefly, wax shavings were added to desert sand (maintained at 75 °C in an open-to-air motorized mixer) in a ~1:500 mass ratio. This forced the wax to melt and form a conformal coating on sand grains during mixing; then, the mixer was turned off, and SHS was used as-is after the mixture had cooled down under ambient conditions (Fig. 1a).

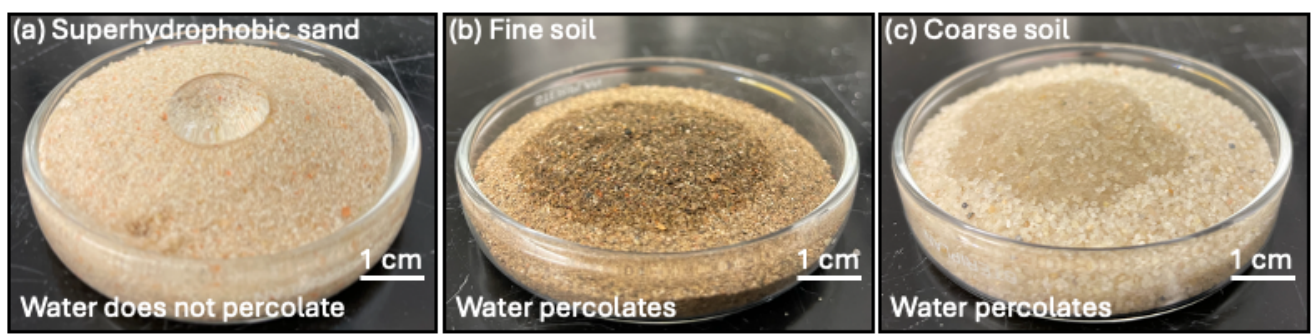

**Figure 1:** Physical appearance and wetting behavior of sand/soil grains studied here: (a) Superhydrophobic Sand (SHS) mulch that prevents intrusion of water; (b) Fine soil absorbs water readily and (c) Coarse soil also absorbs water readily. Note: SHS is lighter than both the sands, and fine sand has a darker tone than coarse sand. A comparison of several physical properties of the sands, relevant to this work, are presented in Table 1.

The efficacy of SHS mulching was interrogated on silica grains as surrogates of sandy soils with the mean particle size distributions of: (i) 219 μm (hereafter referred to as "fine soil") and (ii) 758 μm (hereafter referred to as "coarse soils"). Fig. 1b–c captures the physical appearance of the two soils. Table 1 lists the results of dry sieve analysis (particle size), capillary rise to design experiment vis-à-vis water table depth, photographic albedo estimation, and field capacity measurements. Notably, we did not have thermal conductivity data of these granular materials.

**Table 1.** Relevant physical properties of SHS mulch and representative sandy soils investigated in this study. The reported mean particle size $\mu$ and standard deviation $\sigma$ were obtained by analyzing the size probability distribution function from the sieve analysis. For SHS mulch, particle size is given as the observed range.

| Experimental data | Fine Soil | Coarse Soil | SHS mulch |
|---|---|---|---|
| Particle size (μm) | $\mu = 219, \sigma = 156$ | $\mu = 758\ \sigma = 97$ | Range: 125–710 |
| Density (g/cm$^3$) | 1.68 | 1.62 | 1.5 |
| Capillary rise (cm) | 84 | 8 | 0 (water repellent) |
| Field capacity | 23% | 18% | 0 (water repellent) |
| Color | Brownish | Whitish | Yellowish |
| Albedo | 18.5% ± 2.5% | 40.0% ± 3.7% | 44.1% ± 1.9% |

### 2.2. Laboratory experiments on soil evaporation

To quantify evaporative flux and temperature profiles under controlled irradiation, we constructed a column-based experimental system (Fig. 2a). Twelve polyacrylate columns (45 cm height; 10 × 10 cm² cross-section) were arranged in a 3 × 4 array and exposed to a calibrated overhead lamp array positioned ~50 cm above the surface to ensure uniform top-down heating. A white screen with 10 × 10 cm² apertures confined radiation to the column surfaces and minimized lateral heating effects.

Thermocouples were embedded at the soil surface and at depths of 3, 13, 23, and 43 cm to monitor steady-state temperature profiles in real time. Columns were filled with either fine or coarse soil (Section 2.1), and SHS mulch layers of prescribed thickness were applied where appropriate; mulched and unmulched columns were interspersed to ensure comparable irradiation conditions.

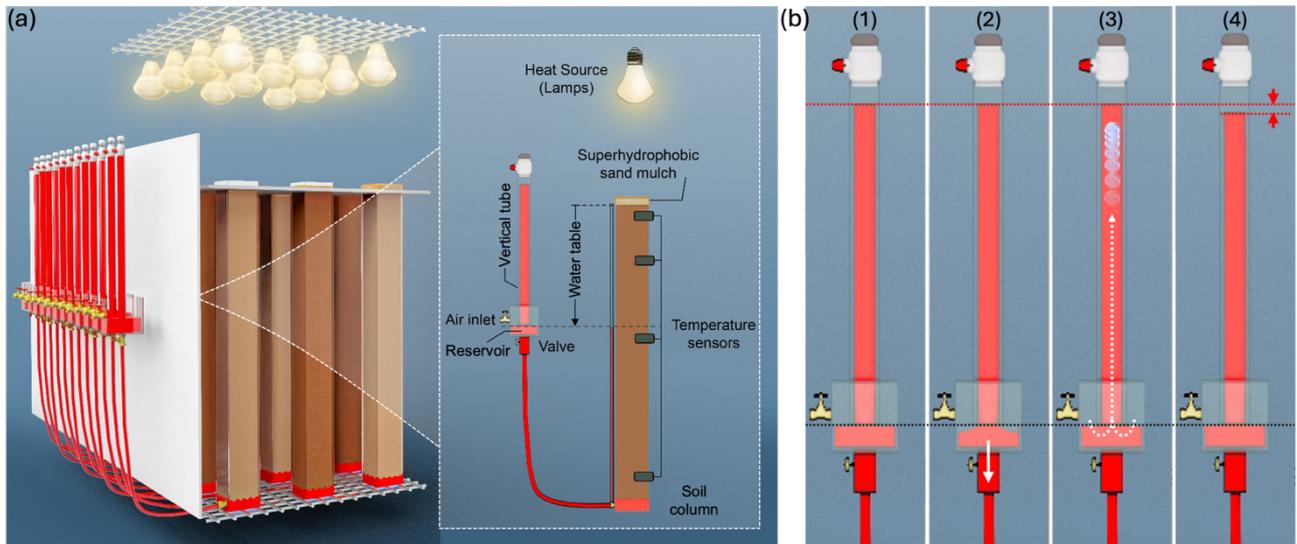

**Figure 2:** Experimental setup. (a) An illustration of the setup for the column evaporation experiments. The inset shows the representative schematic of each of the column. (b) An illustration of the reservoir replenishment process from the vertical tube: (1) Initial condition: the water level in the reservoir is at the water table (indicated by the black dotted line). (2) Reservoir depletion: evaporation from the soil column draws water from the reservoir so that the water level decreases below the water table. (3) Air bubble entrance: an air bubble forms at the tube–reservoir interface and rises upwards. (4) Reservoir replenishment: when the bubbles reach the head space of the vertical tube, the water volume in the vertical tube transfers to the reservoir to compensate for the water loss. Note: the water was dyed red and placed against a white background to facilitate high contrast imaging to aid post-processing.

To maintain hydraulic continuity and quantify evaporation, each column was connected to an atmospheric reservoir system. Water was supplied from the bottom upward to avoid air entrapment. As evaporation depleted the reservoir, water from an inverted vertical tube replenished the reservoir volume, and the corresponding displacement enabled precise measurement of evaporative losses (Fig. 2b).

We define the water table as the vertical distance between the soil surface and the reservoir water level. By systematically varying this distance, we isolated the role of capillary rise in sustaining surface moisture and regulating evaporation.

As evaporation proceeded, reservoir depletion caused a measurable decline in the water level of the connected vertical tube. This displacement was recorded at 5-minute intervals using time-lapse imaging. The dyed water and high-contrast background enabled automated image processing to quantify cumulative evaporative losses over time (ESI Movie 1). From these data, we calculated steady-state evaporative fluxes for saturated soils under different SHS thicknesses and water table depths.

## 3. Results
### 3.1. Soil evaporation flux as a function of SHS mulch thickness and water table

We first examined saturated conditions (water table = 0), representative of the root zone immediately following irrigation. Under steady irradiation, the soil–water–mulch system reached thermal equilibrium while evaporation from the column surface was continuously replenished from the connected reservoir, enabling direct quantification of steady-state flux.

Application of SHS mulch substantially suppressed evaporative losses in both soil types (Fig. 3a). In fine soil, 5 mm and 10 mm SHS layers reduced daily evaporative flux by 65% and 83%, respectively, relative to unmulched controls. In coarse soil, the corresponding reductions were 63% and 70%. Thus, even a thin (5 mm) SHS layer reduced evaporation by approximately two-thirds, while a 10 mm layer reduced flux by up to four-fifths.

The reduction in flux translated directly into prolonged moisture retention (Fig. 3b). For fine soil, 5 mm and 10 mm SHS layers increased the time required to deplete the reservoir by approximately 3× and 6×, respectively. In coarse soil, retention times increased by approximately 2× and 3×. These results demonstrate that SHS mulch markedly extends the persistence of soil moisture under irradiation

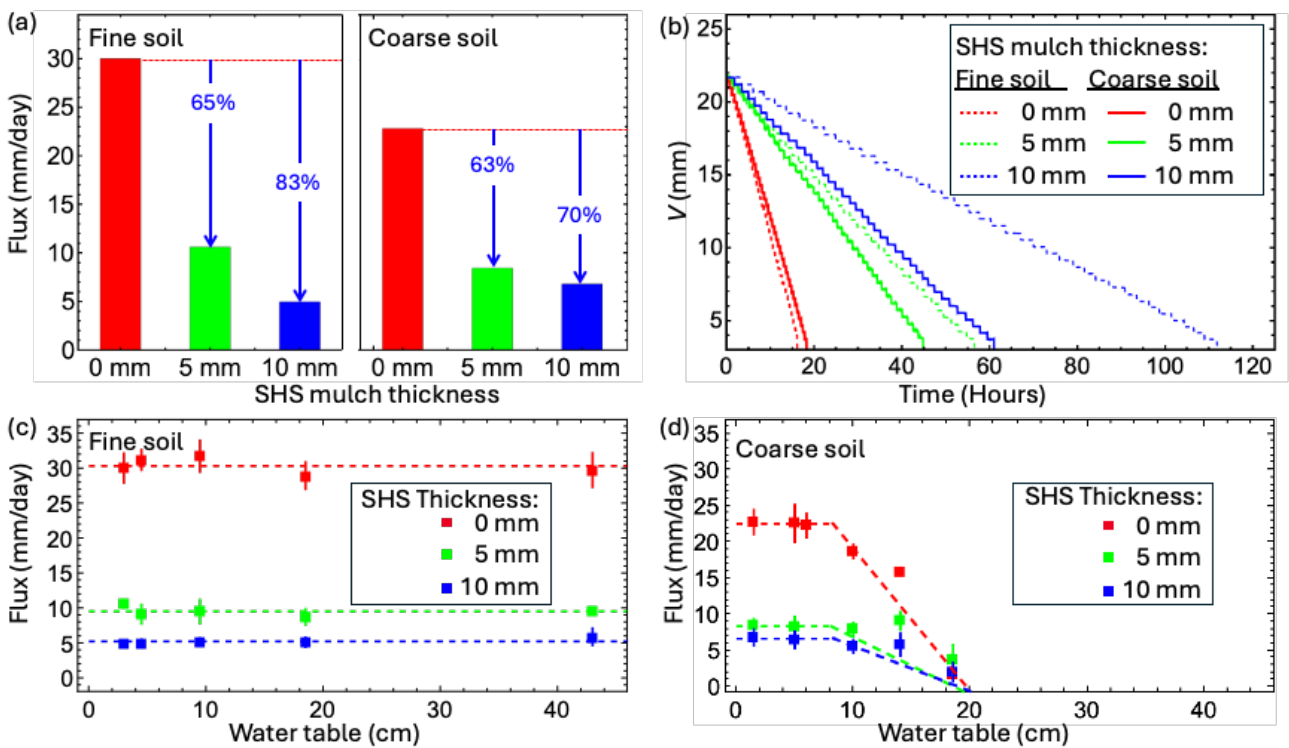

**Figure 3:** Column evaporation experimental results. (a) Evaporative flux for saturated fine and coarse soils under different SHS thicknesses. The results show a significant evaporative flux reduction when SHS mulch is applied. (b) Volume of water in the reservoir as a function of time during the experimental run recorded after the first volume change. Moisture retention increases significantly when the SHS is applied. (c) Evaporative flux as a function of water table of fine and (d) coarse soil with different mulch thicknesses. Note: the flux has been normalized by the cross-sectional area of the column to be presented in mm/day unit.

Here, a question arises: Why does the time-dependence of the evaporative water losses vary quite significantly for the two soils even though the SHS mulch is the same? Before we investigate this matter, we present the effects of water table depth on the evaporation rates. Curiously, for the fine soil, varying the water table from zero to 43 cm did not change the evaporative flux at all (Fig. 3c) – it was 31 mm/day, and SHS mulching reduced it to 10 mm/day (-65%) and 5 mm/day (-83%) at 5 and 10 mm thickness, respectively. In contrast, for the coarse soil, with a capillary rise of 8 cm, evaporative fluxes varied dramatically with the water table depth (Fig. 3d). When the water table is deeper than 8 cm, evaporative fluxes from unmulched soil are reduced from 22.5 mm/day to 8.5 mm/day (-63%) and 6.5 mm/day (-70%) for 5 mm and 10 mm thickness. We explain these findings based on capillary rise of water, which is the height of water above water table at which the surface tension and the weight forces are at balance. Based on experimental observation, the fine soil presented a capillary rise of $l_c \approx 84$ cm (Table 1), whereas the coarse soil had a capillary rise of 8 cm. Thus, in the former case, water was able to climb to the top of the column even if the water table was at its lowest level (45 cm); on the other hand, for the coarse soil, as the water table was set > 8 cm, capillarity could not draw the water up to wet the topsoil. In the latter case, therefore, the dry soil began to act as a mulch and the evaporative flux sharply decreased for both mulched and unmulched scenarios and ultimately becoming undetectable by our technique. We close this section by noting that capillary rise $l_c$ for a given liquid scales with the average pore size, $r$, as

$$l_c \propto 1/r, \qquad (1)$$

and the particle size distributions in Table 1 provide a semi-quantitative basis for this observation.

### 3.2. Soil temperature profiles

To elucidate the differences in steady-state evaporation between fine and coarse soils, we measured their temperature distributions under identical irradiation (Fig. 4). In the unmulched condition, fine soil reached a surface temperature of 41 °C, whereas coarse soil reached 38 °C. The corresponding evaporative fluxes were 33 mm/day and 24 mm/day, respectively—a 37.5% difference.

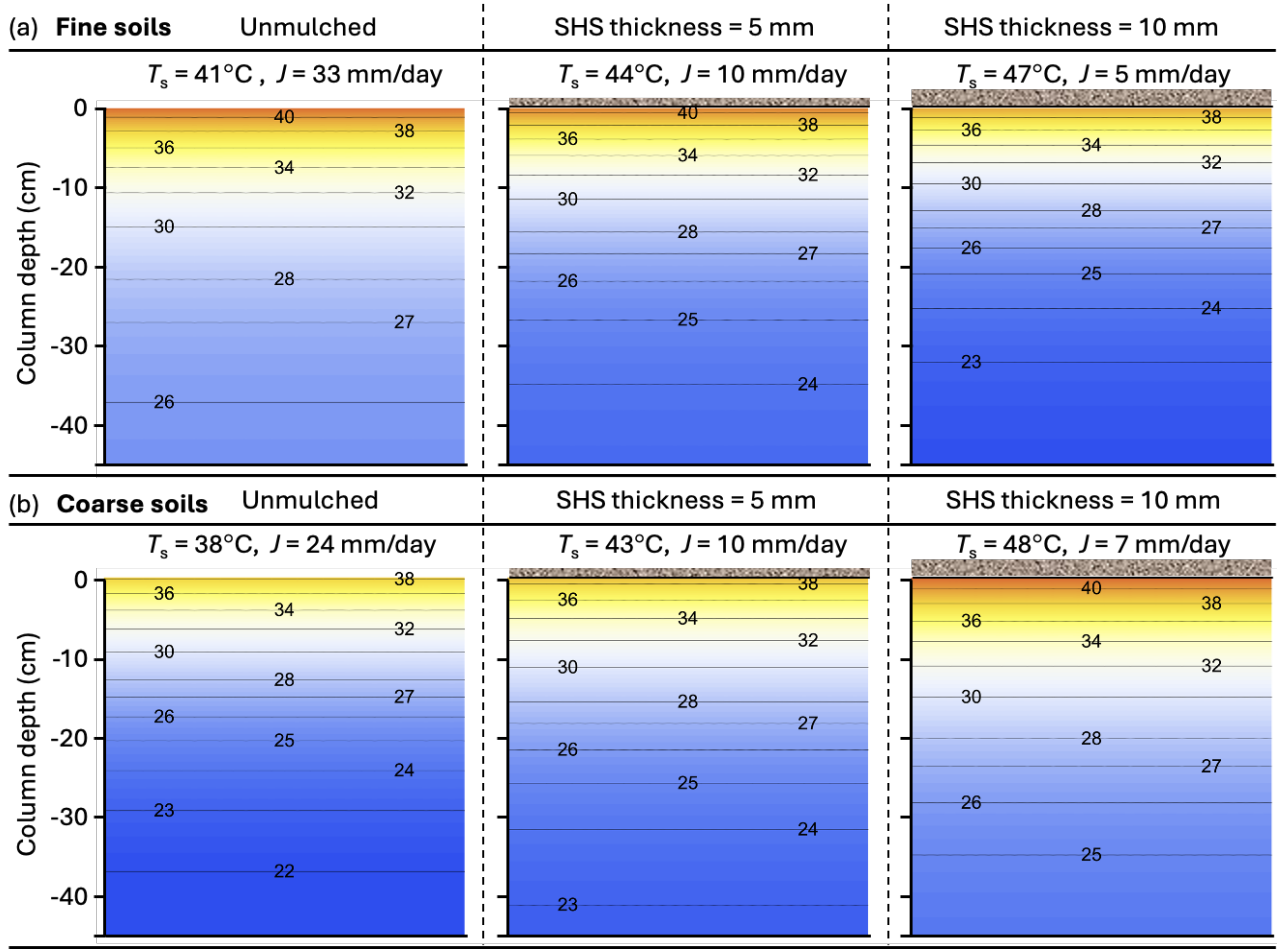

**Figure 4:** Experimentally measured temperature profiles of mulched and unmulched soil columns (fine and coarse). Representative temperature profiles of (a) Fine and (b) Coarse soil columns exposed to uniform radiation, with varied mulch thicknesses and a constant 5 cm water table depth at steady state. The contours show the temperature at the certain column depth. $T_s$ and $J$ are the soil or mulch surface temperature and the evaporative flux respectively. The gray rectangles above middle and right panels are to illustrate the thickness of the SHS mulch, which is not to scale.

This disparity is consistent with differences in surface albedo. Wet fine soil appears darker than coarse soil (Table 1; Fig. 1), absorbs a greater fraction of incident radiation, and consequently attains a higher surface temperature. Since evaporation in saturated soils is strongly temperature-dependent, the higher flux from fine soil follows directly from its elevated surface temperature.

After SHS application, however, the thermal behavior deviated from this albedo-based expectation. Because the mulch surface color is similar in both cases, one might anticipate comparable interfacial temperatures. Instead, contrasting trends emerged. For fine soil, the SHS–soil interfacial temperature decreased slightly with increasing mulch thickness (41 °C at 0 mm to 39 °C at 10 mm). In contrast, coarse soil exhibited an increase in interfacial temperature (38 °C at 0 mm to 41 °C at 10 mm). Temperature elevation was also more pronounced at deeper depths in the coarse soil column.

At the mulch–air interface, temperature increased monotonically with SHS thickness in both soils. For fine soil, surface temperature rose from 41 °C (unmulched) to 44 °C (5 mm) and 47 °C (10 mm). For coarse soil, it increased from 38 °C to 43 °C and 48 °C, respectively.

These observations suggest that SHS mulch modifies not only surface radiation balance but also subsurface heat transport. Specifically, they raise three mechanistic questions addressed in the following section through heat-transfer modeling:

1. Why do mulched fine and coarse soils exhibit different interfacial temperature trends?
2. Why does the SHS mulch surface become hotter than the unmulched saturated soil?
3. How do these temperature profiles interact with mulch thickness to regulate evaporative flux?

### 3.3. Mathematical modeling of heat transfer

To interpret the observed temperature profiles, we developed a first-principles heat-transfer model accounting for radiation, conduction, convection, and latent heat removal due to evaporation[26]. At steady state, the energy balance at the soil or mulch surface includes:

Radiative exchange:
$$Q_r = \varepsilon \sigma A_c (T_s^4 - T_a^4), \quad (2)$$

Conduction into the soil column:
$$Q_c = -\kappa A_c (\Delta T / \Delta z), \quad (3)$$

Convective heat exchange with air:
$$Q_h = h A_p (T - T_a), \quad (4)$$

Latent heat associated with evaporation:
$$Q_e = A_c J \delta t L. \quad (5)$$

Here, $\varepsilon$, $\sigma$, $\kappa$, $h$, $A_c$, $\delta A_p$, $J$, $\delta t$, $L$, $T_s$, $T_a$, $T$, and $(\Delta T/\Delta z)$ are, respectively, the emissivity, Stefan–Boltzmann constant, heat conductivity, convective heat transfer coefficient, cross-sectional surface area, perimeter along $A_c$, evaporative flux, timestep, latent heat of vaporization, steady-state soil or mulch surface temperature, ambient temperature, soil layer temperature, and the gradient of the soil temperature respectively. Soil columns were discretized into layers and each layer would absorb and transfer heat at every timestep, and the temperature of each layer was updated by $\Delta T = Q/(m C_s)$, where $Q$, $m$, and $C_s$ are the net heat energy, mass, and heat capacity of the soil layer respectively. The details of how the model is implemented into a simulation code is provided in the Methods section and ESI.

We benchmarked this model by first comparing the relative contributions of the heat transfer modes in the light of the experimental details. Our analysis revealed that the temperature profile of the soil columns at steady state was most sensitively dependent on the soil thermal conductivity, $\kappa$ (ESI Fig. S2). Other parameters such as $h$ and $C_s$ did not significantly impact the temperature profile. This situation is analogous to the classical case of a heat exchanger fin (or extended surface) with the steady state temperature profile prescribed by the formula [26]

$$T(z) = T_a + (T_{sm} - T_a)e^{zM}, \quad (6)$$

where $M = \sqrt{(hA_p)/(\kappa A_c)}$ is a decay parameter (related to the Biot number and geometry) and $T_{sm} = T_s$ for the unmulched case. Equipped with this insight, we fitted Eq. (6) into the experimental data presented in Fig. 4 to obtain the values of $M$ for the 6 cases. Interestingly, we found that the obtained the $M$ values are specific for the soil type and were independent of the mulch thickness: $8.1 \pm 0.1$ m$^{-1}$ for the fine soil and $6.6 \pm 0.1$ m$^{-1}$ for the coarse soil. In addition, since $h$, $A_p$, and $A_c$ were about the same for all the columns, we deduced that the coarse soil has a higher heat conductivity than the fine soil, i.e., $\kappa_{CS} > \kappa_{FS}$.

The effect of $\kappa$ is not immediately visible from Fig. 4 as every case has a different temperature profile. However, when the temperature at depth $z$ is expressed as non-dimensional temperature, defined as $T^* = \frac{(T(z) - T_a)}{(T_{sm} - T_a)}$, the temperature profile depends only on the type of the soil, regardless of the mulch thickness. To confirm this, we used the $\kappa_{FS}$ and $\kappa_{CS}$ values (obtained above) as input parameters for simulations. The simulated steady state temperature profile and its comparison with experimental results are presented in Fig. 5a-b. This simple numerical approach successfully captures the temperature profiles in our experiments.

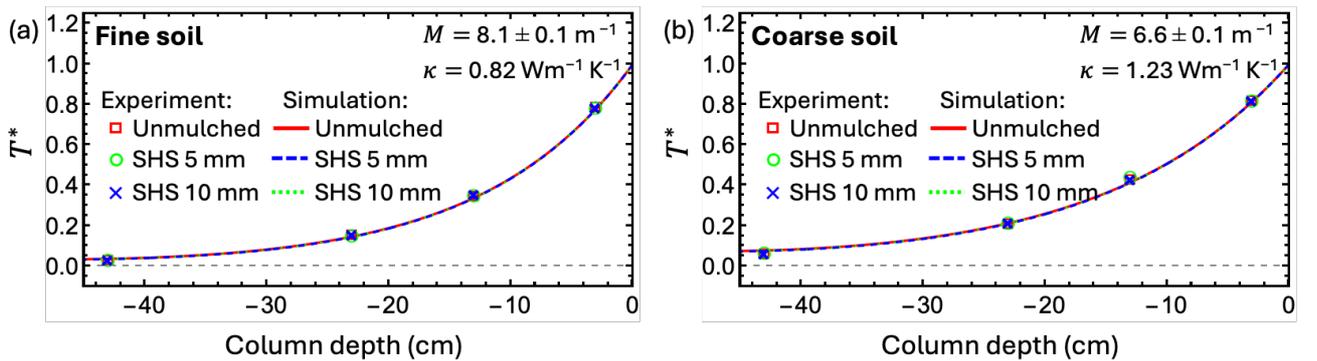

**Figure 5:** Simulation results. Non-dimensional temperature profiles of (a) Fine soil and (b) Coarse soils. The gridline at $T^* = 0$ is a guide to compare between the two panels.

The $\kappa$ value also critically influenced the temperature of the mulch. Although the temperature profile inside the SHS mulch was not experimentally recorded, it is intuitive to expect that the $\kappa$ value of dry and porous SHS mulch is lower than that of the wet soil[27]. We estimated the $\kappa$ value of the SHS mulch by measuring the temperature profile of a dry sand column, yielding $\kappa_{SHS} = 0.37 \pm 0.03$ W-m$^{-1}$K$^{-1}$. We postulated that this low $\kappa$ value causes heat to accumulate within the SHS mulch – therefore, the thicker the mulch, the longer it takes for the heat to be conducted to the wet soil column underneath. As a result, the top mulch temperature $T_{ma}$ increases with the thickness of the SHS mulch, as observed in experiment. We applied our simulation tool to probe the effect of SHS mulch thickness on $T_{ma}$ under constant radiation (Fig. 6) and found reasonable agreement with the experiments.

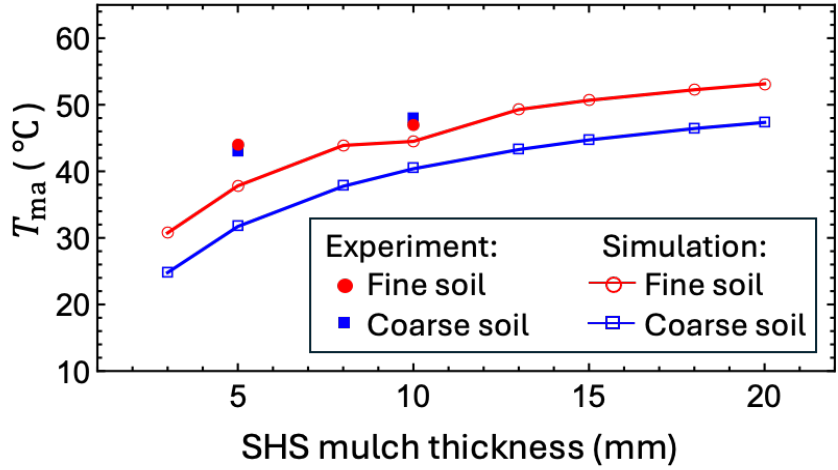

**Figure 6:** The comparison of experiment and simulation values of the top mulch temperature $T_{ma}$ as function of the SHS mulch thickness.

We next examined the governing mechanism of evaporative flux in the unmulched configuration. For saturated soils directly exposed to air, evaporation is modeled using the Hertz–Knudsen relation derived from kinetic gas theory[28]:

$$J_u = \frac{\alpha_b}{\sqrt{R}} \left( \frac{p_T}{\sqrt{T_T}} - \frac{p_a}{\sqrt{T_a}} \right), \quad (7)$$

where $\alpha_b$, $R$, $T_T$, $p_T$, and $p_a$ are respectively the accommodation parameter, gas constant for water vapor (461.52 JkgK$^{-1}$), soil surface temperature, vapor pressure evaluated at the soil surface temperature, and at ambient temperature. Because ambient conditions were held constant, Eq. (7) predicts that evaporative flux depends solely on the surface temperature $T_T$. Thus, irrespective of soil type, the unmulched evaporation rate should be controlled primarily by surface temperature. To test this prediction, we plotted experimentally measured fluxes for both fine and coarse soil s as a function of $T_T$ (Fig. 7a). Remarkably, all data collapse onto a single master curve consistent with the Hertz–Knudsen relation. This collapse confirms that, under saturated and unmulched conditions, evaporation is governed by surface temperature rather than soil texture or subsurface thermal properties.

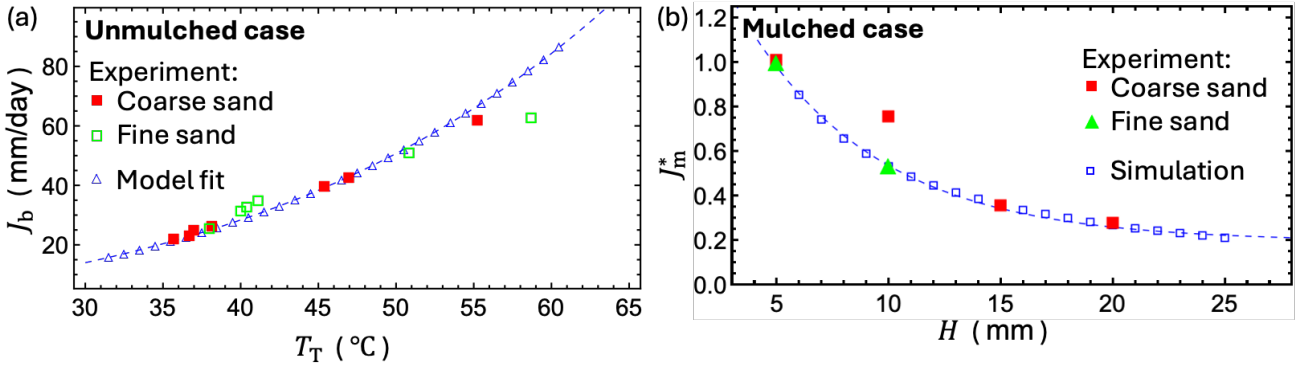

**Figure 7:** (a) Comparison of model fit and experimental evaporative flux for the unmulched fine and coarse soil columns. (b) The comparison of experimental and the simulated flux for the mulched cases. The flux values were normalized by the flux of 5 mm mulch thickness ($J_m^* = J_m/J_{mH5}$).

In the presence of SHS mulch, direct liquid–air contact is prevented and evaporation proceeds through vapor diffusion across the porous mulch layer. The evaporative flux is therefore modeled using Fick's law[29]:

$$J_m = \phi^{4/3} D(\rho_{sm} - \rho_{ma})/H, \qquad (8)$$

where, $J_m$, $\phi$, $D$, $H$, $\rho_{sm}$ and $\rho_{ma}$ are the evaporative flux in the presence of mulch, porosity of superhydrophobic sand mulch ($\phi$ = 0.416), the diffusion coefficient of water, mulch thickness, water vapor density at the soil-mulch, and at mulch-air interfaces respectively.

In contrast to the unmulched case, flux in the mulched configuration depends explicitly on the diffusion path length $H$ and the vapor concentration gradient across the mulch. This formulation predicts an inverse dependence of flux on mulch thickness, characteristic of diffusion-limited transport.

To test this scaling, we compared experimental and simulated fluxes across both soil types (Fig. 7b). Normalizing flux by its value at $H = 5$mm, $J_m^* = J_m/J_{mH5}$, reveals consistent collapse of the data onto the diffusion-controlled trend. The agreement between model and experiment confirms that SHS mulch shifts evaporation from a surface-temperature-controlled regime to a diffusion-limited regime governed by vapor transport resistance.

## 4. Discussion

The combined experimental and modeling results demonstrate that SHS mulch induces a transport-regime transition in soil evaporation. In the unmulched configuration, evaporative flux collapses onto the Hertz–Knudsen relation when plotted against surface temperature, confirming that evaporation from saturated soils is surface-temperature controlled. Differences between fine and coarse soils arise primarily from albedo and thermal conductivity effects on steady-state surface temperature (41 °C vs 38 °C), leading to corresponding flux differences (33 vs 24 mm/day).

Application of SHS fundamentally alters this mechanism. By eliminating liquid–air contact and introducing a porous diffusion barrier, evaporation becomes governed by Fickian vapor transport across a finite thickness $H$, with $J_m \propto 1/H$. Experimental data normalized by the 5 mm case collapse onto the diffusion-scaling prediction, confirming diffusion-limited behavior. The reduction in flux (up to 83% for 10 mm SHS) and the corresponding 2–6× increase in moisture retention time follow directly from the added vapor transport resistance.

The reversal in relative evaporation rates between fine and coarse soils under mulched conditions is explained by differences in thermal conductivity. In the temperature-controlled regime, darker fine soil reaches higher surface temperatures and evaporates more rapidly. In the diffusion-limited regime, however, the vapor density gradient across the mulch depends on the soil–mulch interfacial temperature, which is governed by subsurface heat transport. Because coarse soil exhibits higher thermal conductivity ($\kappa_{CS} > \kappa_{FS}$), it sustains higher interfacial temperatures under identical irradiation, leading to slightly larger vapor density differences and therefore higher flux relative to fine soil.

The low thermal conductivity of SHS ($\kappa_{SHS} \approx 0.37$ W·m$^{-1}$K$^{-1}$) increases thermal resistance at the surface, causing mulch–air interface temperature to rise with thickness. However, simulations show that subsurface temperature profiles remain dominated by soil thermal conductivity rather than mulch heating. Thus, evaporation suppression arises primarily from diffusion resistance rather than surface cooling.

Water-table experiments further confirm that capillary connectivity determines whether evaporation is supply-limited or transport-limited. Fine soil (capillary rise ≈ 84 cm) maintains connectivity across the column height, whereas coarse soil (≈ 8 cm) becomes self-mulching once the water table exceeds its capillary rise. SHS therefore provides the greatest relative benefit in capillary connected systems where surface moisture would otherwise sustain high evaporative losses.

The present study focuses on saturated, steady-state conditions. In field settings, soils undergo daily wetting–drying cycles under irrigation. As shown in Fig. 3d, once the upper soil layer dries, it introduces intrinsic diffusion resistance and partially mimics the effect of SHS. Quantifying SHS performance under transient moisture conditions therefore requires time-resolved gravimetric or field-scale measurements. In addition, synergistic interactions with soil amendments that modify thermal conductivity, albedo, and hydraulic conductivity, such as engineered biochar[25], warrant systematic investigation. Finally, the current experiments do not account for canopy shading, which would likely reduce surface heating in SHS-treated soils due to enhanced moisture retention and vegetative growth. Consequently, the observed surface temperature increases may represent an upper bound.

Overall, SHS suppresses evaporation by imposing a controllable vapor diffusion barrier, even under elevated surface temperatures. The magnitude of suppression is governed by mulch thickness, porosity, vapor diffusivity, and soil thermal conductivity, providing a quantitative framework for optimizing SHS performance across soil textures and irradiation conditions.

## 5. Conclusions

This study demonstrates that Superhydrophobic Sand (SHS) mulch substantially suppresses evaporative water loss from saturated sandy soils by altering the governing transport mechanism. Under controlled irradiation, 5 mm and 10 mm SHS layers reduced evaporative flux by up to 65% and 83%, respectively, and extended moisture retention times by factors of approximately 2–6 relative to unmulched soils. Mechanistically, SHS shifts evaporation from a surface-temperature-controlled regime to a diffusion-limited regime governed by vapor transport resistance through the porous mulch layer. While increasing mulch thickness elevates the mulch surface temperature due to its low thermal conductivity, subsurface temperature profiles remain primarily determined by the soil's intrinsic thermal conductivity. In saturated conditions, coarse soil —with higher thermal conductivity— exhibits cooler subsurface temperatures than fine soil. Together, the experimental measurements and heat–mass transfer modeling clarify the coupled roles of albedo, thermal conductivity, capillarity, and vapor diffusion in regulating evaporation. These insights provide a physical basis for optimizing SHS thickness and soil selection to enhance irrigation efficiency in arid agricultural and landscaping systems.

**Open Research Section**

Experimental and simulation data sets in addition to all the codes used for this research are publicly available at: https://github.com/amrhz1/SandX.git

**Acknowledgments**

HM acknowledges KAUST for funding (Grant#: BAS/1/1070-01-01). AGJ acknowledges meaningful discussions with Prof. Carlos J. Santamarina (KAUST). The co-authors thank Mr. Yinfeng Xu (KAUST) for his assistance in quantifying the radiation intensity of the light source. AAZ acknowledges Mr. John Peresson (KAUST) for assistance in experimental work during his sick leave. LE acknowledges Edelberto Manalastas for assistance in building the experimental setup. The co-authors thank Dr. Gloria Fuentes, Dr. Ana Bigio, and Dr. Thom Leach, Scientific Illustrators at KAUST, for preparing Fig. 2.

**Author Contributions**

HM and AGJ co-invented SHS and conceptualized the study. AGJ designed and built the first generation of the experimental setup and the results of this study were reported in this PhD thesis (https://repository.kaust.edu.sa/items/51909043-878b-43c7-a47d-a7904d8c7865). The next generation of experimental setup was designed and built by LE. AAZ contributed to experimental design, regular maintenance, and leading the comprehensive set of experiments and data analysis presented in this report. MSS developed the mathematical model, performed the simulations and analyzed simulation results. AAZ and MSS wrote the first draft of the manuscript, which HM substantially edited. JZ wrote the Python script for data acquisition.

# Supplementary Information

**Contents of this file**

S1: Preparation of superhydrophobic sand mulch

S2: Particle size analysis

S3: Experimental setup and procedures

S4: Albedo Measurement

S5: Numerical method

S6: Data and Codes

Additional Supporting Information (Files uploaded separately)

ESI Movie 1: A video showing the time lapse of the water volume change in the vertical tube during experimental run. Video analysis was performed as a probe of evaporative flux measurement.

## Introduction

This supporting information provides additional details on the experimental and numerical methods used in this study. It includes descriptions of the preparation of the superhydrophobic sand mulch, the experimental setup and procedures, and the technique used for albedo measurement. A section outlining the numerical modeling approach is also provided.

All experimental data and model outputs referenced in the main manuscript are documented here, and all relevant codes and datasets are made available through an external repository via a permanent link. These materials are intended to enhance transparency and reproducibility.

## S1 Preparation of superhydrophobic sand mulch

SHS was produced by mixing desert sand with 0.5% of paraffin wax by mass in a 1.5 L round flask. The flask was tilted and rotated using a rotary evaporator at a rate of ~ 35 RPM, at a temperature equal to 75 °C.

## S2 Particle size analysis

The particle size characterization was performed using sieve analysis, from which we obtain the cumulative particle passing for each sieve size. The log-normal cumulative distribution function (CDF)

$$D(x) = \frac{1}{2}\left[1 + \mathrm{erf}\left(\frac{\ln x - \beta}{\alpha\sqrt{2}}\right)\right], \tag{S1}$$

was then fitted to the sieve analysis data to obtain the distribution function parameters $\beta$ and $\alpha$, as shown in Fig. S1.

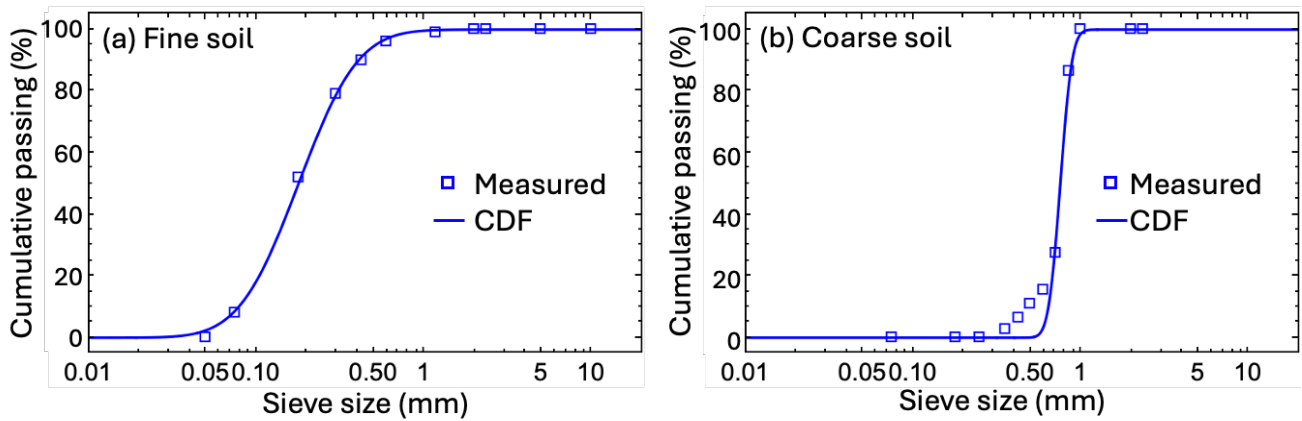

**Fig. S1:** Sieve analysis of (a) fine and (b) coarse soils and their fit with the cumulative distribution function to obtain the parameters of the probability distribution function.

Here, $x$ is the sieve size. The parameters were then used to construct the log-normal probability distribution function

$$P(x) = \frac{1}{\alpha x \sqrt{2\pi}} e^{-(\ln x - \beta)^2/(2\alpha^2)} . \qquad (S2)$$

The mean particle size $\mu$ and the standard deviation $\sigma$ were the calculated from the moments of the probability distribution function as $\mu = \int_0^\infty x P(x)\, dx$, and $\sigma^2 = \int_0^\infty x^2 P(x)\, dx$.

## S3 Experimental setup and procedures

Capillary rise experiment was conducted using a 1 m long tube by 1 cm diameter. The tube was filled with sandy soil and fixed vertically while having a water source at the bottom and left for a sufficient time to see the maximum rise while the meniscus is still visible.

As for the field capacity jars with holes from the bottom were filled with soil then completely submerged in water for a day to make sure the soil was fully saturated. The jars were then removed from the water and covered from the top and the water was allowed to drain for 2 days, the percent increase in mass is the field capacity.

During experimental run, heat energy was supplied through radiation using a mesh containing twenty-six 100-watt incandescent bulbs. The columns were made from acrylic with the following dimensions: 10 cm (length) × 10 cm (width) × 45 cm (height). This set-up only allowed heat to be applied from the top of the soil columns, thus, not allowing any heat from the side of the column to interfere with the evaporation as an extra energy provider. Temperature sensors are placed at depths of 3 cm, 13 cm, 23 cm, and 43 cm, and an air temperature and humidity sensor are positioned 1 cm above the column surface.

## S4 Albedo Measurement

Images were taken at different positions with different light intensities for fine, coarse and super hydrophobic sand above a white paper. Using the ImageJ software the intensities of light at each sample and the white paper was recorded. Assuming that the albedo of the white paper was equal to 0.6 and $I_x$ is the white paper intensity, and $I_s$ is the sample intensity, the albedo was measured as Albedo = 0.6 ($I_s/I_x$).

## S5 Numerical method

**Algorithm.** The column evaporation experiment can be mathematically modeled using a quasistatic approach, where equilibrium is assumed to be satisfied at any time. The soil column was discretized into layers. At every time step $\delta t$, each layer experiences a temperature change depending on the total heat energy change $Q_T = Q_{in} - Q_{out}$. For the top layer, $Q_{in}$ is the heat energy source, and $Q_{out}$ is the sum of the heat losses due to radiation, conduction, and convection, as given in Eqs. (2-4). $Q_{in}$ is given by $Q_{in} = P_s A_c \delta t$, where $P_s$ and $A_c$ are radiation pressure and cross-sectional area. In addition, for the bare case, heat loss due to evaporation is calculated as $Q_e = A_c J \delta t L$, where $L$ is the latent heat of vaporization and $J_u$ is the evaporation rate, as given in Eq. (7). Teten's equation

$$p = 0.61078 \ \text{Exp}\left(\frac{17.27T}{T+237.3}\right), \tag{S3}$$

is used to calculate the vapor pressures ($p_T$ and $p_a$) in Eq. (7) of the main text. The temperature change is calculated as $\Delta T = Q_T/(mc)$, where $m$ and $c$ are the mass and heat capacity of the corresponding soil layer, respectively. After updating the temperature of the top soil layer, the temperature of the lower soil layer is calculated using the conducted heat from the upper layer as $Q_{in}$, and the sum of conduction and convection heat loss only as $Q_{out}$. For the mulched case, evaporation does not occur at the top layer, but instead at the soil layer under the mulch. In this case, the evaporation rate of the mulched case $J_m$ is given by Eq. (8) of main text where the value of the vapor densities ($\rho_{sm}$ and $\rho_{ma}$) as a function of soil temperature is obtained from empirical relation

$$\rho = 6335 + 0.6718T - 2.0887 \times 10^{-2}T^2 + 7.3095 \times 10^{-4} \times T^3. \tag{S4}$$

The calculations were performed iteratively until thermal equilibrium was reached, defined as the condition where $Q_{in} \cong Q_{out}$.

This algorithm was implemented into a simulation code using *Mathematica* software and the code is made available and can be accessed from the link provided.

**Input parameters.** The input parameter includes heat conductivity $\kappa$ of soil and SHS mulch, density $\rho$, heat capacity $C_s$, heat transfer coefficient $h$, intensity of radiation source, mulch thickness, albedo, porosity od SHS mulch $\phi$, accommodation parameter $\alpha_b$, simulation dimensions, and time of simulation run $t$. In addition to those, the following constants were used: emissivity $\varepsilon$ = 0.95, Stefan-Boltzmann constant $\sigma$ = 5.67 ×10$^{-8}$ Jt$^{-1}$m$^{-2}$K$^{-1}$, gas constant $R$ = 461.52 Jkg$^{-1}$K$^{-1}$, mass of water molecule $m_w$ = 2.988×10$^{-26}$ kg, and diffusivity of vapor in air $D$ = 2.5×10$^{-5}$ m$^2$s$^{-1}$.

**Parameter analysis.** Fig. S2 shows the temperature response of the system for different scenarios. Fig. S1a shows the temperature profiles for different situations where a specific heat transfer mode is incorporated. We can see from Fig. S2a that when there is no heat transfer, only the top soil temperature increases after absorbing the heat energy. We can also see that heat loss due to radiation alone is not enough to bring the top soil temperature down. It is only when conduction is incorporated that the heat is propagated to the lower part of the soil column. On the other hand, convective heat loss does not seem to contribute significantly.

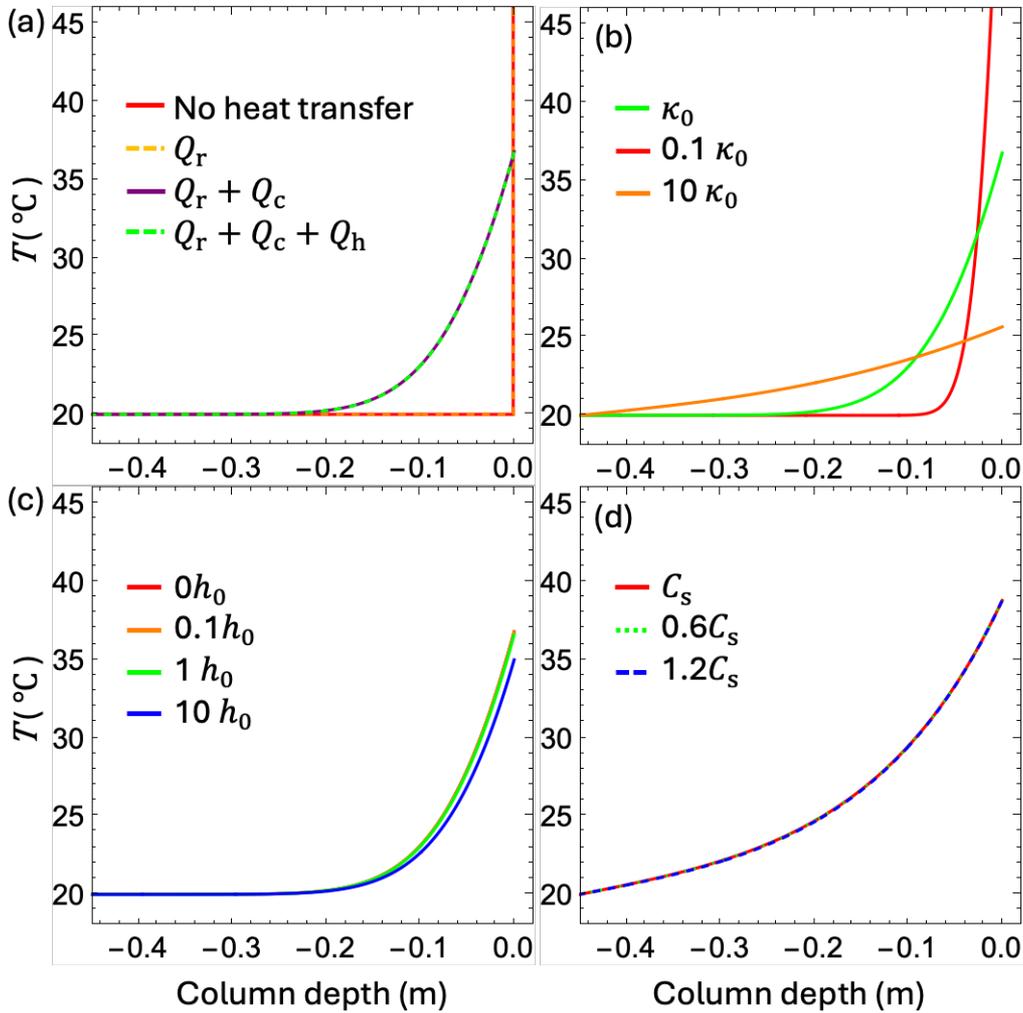

**Fig. S2. Temperature response of the system for different parameter values.** (a) The effect of heat transfer modes to the temperature profile. (b) Temperature profiles for different values of heat conductivity and (c) convective heat transfer coefficient. The simulations for panels (a)-(c) were performed for up to $t = 1$ hour and we have used $\kappa_0 = 3.72$ W m$^{-1}$K$^{-1}$ and $h_0 = 0.4$ W m$^{-2}$K$^{-1}$. (d) Temperature profiles for different values of heat capacities, $C_s = 1600$ J kg$^{-1}$K$^{-1}$. The simulations for panel (d) were performed for up to $t = 48$ hours and the system has reach thermal equilibrium.

We also tested the effect of $\kappa$, $h$, and $C_s$ when all heat transfer modes are incorporated, respectively shown in Fig. S2b, c, and d. We can see that $\kappa$ affect the temperature profiles significantly as different value of $\kappa$ resulted in significantly different temperature profiles. On the other hand, different value of $h$ do not affect the temperature profile so much. Similarly, at thermal equilibrium, $C_s$ does not seem to have any effect to the temperature profile of the soil column. This is analogous to a fin (heat exchanger) whose the temperature profile is given by

$$T(z) = T_a + (T_{sm} - T_a)e^{zM}. \tag{S5}$$

Here, we can see that $C_s$ does not explicitly determine the temperature profile. The $M$ values for the fine and coarse soils can be easily determined by fitting the experimental results with Eq. (S5). The obtained $M$ values then can be used as input parameters for the simulations. The comparison of simulations and experimental results, as well as model fit of Eq. (S5) is provide in Fig. S3 where we can see good the agreement among them.

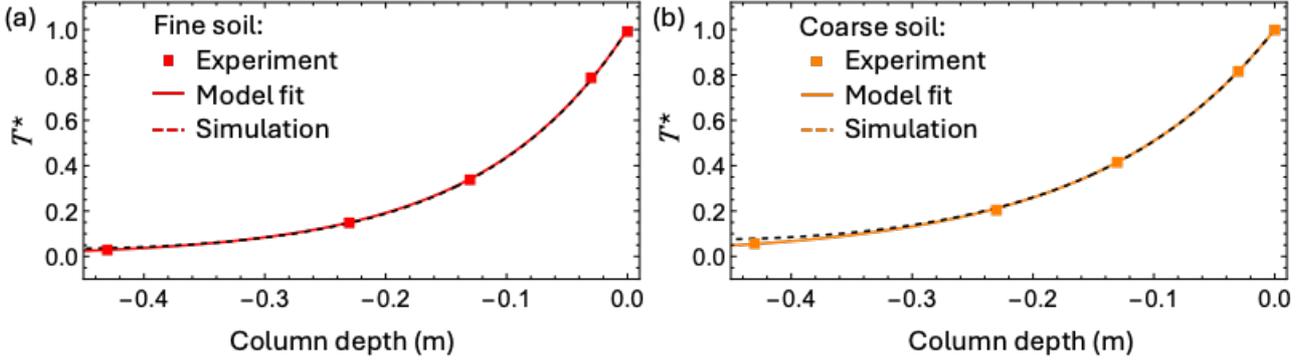

**Fig. S3. Temperature profiles.** Comparison of the temperature profiles for (a) the fine and (b) the coarse soil columns obtained from experimental and simulation.

**Simulation temperature profile.** The simulated temperature profile, soil or mulch surface temperature, and evaporation rate for different column evaporation cases are presented in Fig. S4. The results show a good agreement with the experimental result: the unmulched fine soil has higher surface temperature than the unmulched coarse soil, the lower column temperature of the coarse soil is higher than that of the fine soil, and the surface temperature increases with the mulch thickness. Some difference is also observed. For example, in experiment the lower column temperature is varying from one column to another, whereas in simulation it is rather similar from one to another. This discrepancy is because the ambient temperature in simulation is fixed at 20 °C, while in experiment, the temperature around the soil column can be in thermal equilibrium with the column temperature. However, when the temperature is expressed as non-dimensional temperature, both simulation and experiment temperature profiles are in excellent agreement, as shown in Fig. 5 in the main text.

An interesting future work to improve the agreement between simulation and experiment would be to add insulation layers around the soil columns for the experiment to prevent heat dissipation. In this case, the convective heat dissipation mode also needs to be deactivated in simulation to mimic the experimental condition.

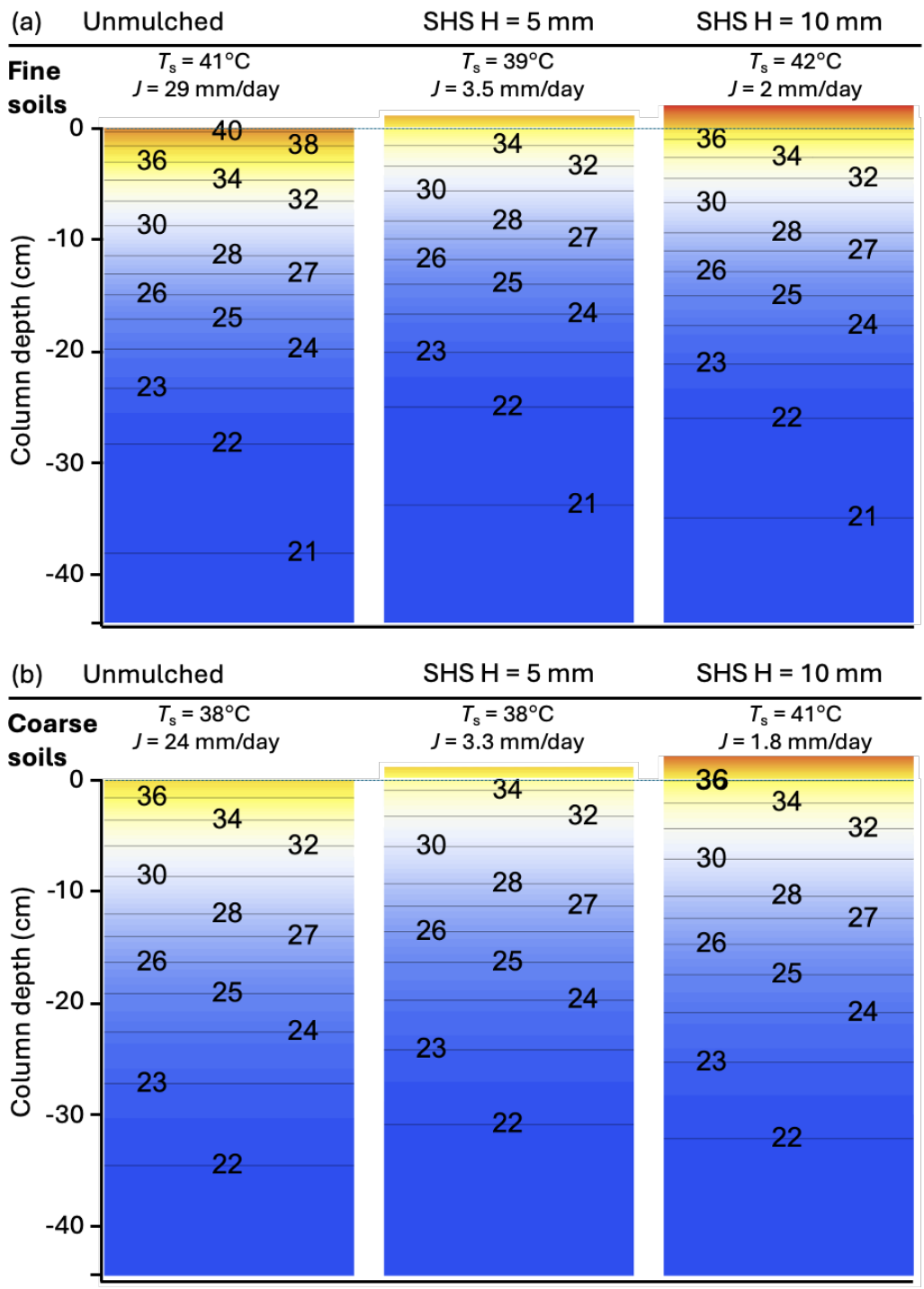

**Fig. S4.** The simulated temperature profiles, surface temperature, and evaporative flux of (a) the fine and (b) the coarse soil columns.

**S6 Data & Codes**

All the data and codes used can be accessed through this link: https://github.com/amrhz1/SandX.git